\documentclass[10pt]{article}
\usepackage{fullpage,times}
\usepackage{amsmath}
\usepackage{amsfonts}

\newtheorem{theorem}{Theorem}

\newtheorem{lemma}[theorem]{Lemma}

\newtheorem{proposition}[theorem]{Proposition}

\newcommand{\qed}{\rule{7pt}{7pt}}
\newenvironment{proof}{\noindent{\bf Proof}\hspace*{1em}}{\qed\bigskip}

\newcommand\be{\begin{equation}}
\newcommand\bea{\begin{eqnarray}}
\newcommand\eea{\end{eqnarray}}
\newcommand\ben{\begin{eqnarray*}}
\newcommand\een{\end{eqnarray*}}

\newcommand{\Ai}{\mathcal{A}}
\newcommand{\LPS}{\mathcal{G}}

\newcommand\n{\nonumber}

\begin{document}

\title{Efficient Discrete Approximations \\ of Quantum Gates}

\author{ Aram W. Harrow \\
~~aram@mit.edu~~\\	
MIT Media Laboratory \\ 20 Ames Street \\ Cambridge, MA 02139
\and
 Benjamin Recht \\
brecht@media.mit.edu\\
MIT Media Laboratory \\ 20 Ames Street \\ Cambridge, MA 02139
\and
 Isaac L. Chuang\\
ichuang@media.mit.edu\\
MIT Media Laboratory \\ 20 Ames Street \\ Cambridge, MA 02139
}

\maketitle

\begin{abstract}
Quantum compiling addresses the problem of approximating an arbitrary
quantum gate with a string of gates drawn from a particular finite
set.  It has been shown that this is possible for almost all choices
of base sets and furthermore that the number of gates required for
precision $\epsilon$ is only polynomial in $\log 1/\epsilon$.  Here we
prove that using certain sets of base gates quantum compiling requires
a string length that is linear in $\log 1/\epsilon$, a result which
matches the lower bound from counting volume up to constant factor.

\end{abstract}
%\pacs{03.67.Lx}

%\terms{quantum computation, physics of computation, optimal compilation}

%\keywords{quantum computation, universal quantum gates, quantum
%circuits, quantum algorithms, quantum compiling}

%%%%%%%%%%%%%%%%%%%%%%%%%%%%%%%%%%%%%%%%%%%%%%%%%%%%%%%%%%%%%%%%%%%%%%%%%%%%%
\section{Introduction}

Quantum computation generalizes computer science to utilize novel
quantum physical resources as elementary building blocks for
information processing~\cite{Bennett00a, Bennett98a,
rieffel:polak:review, Aharonov99b}.  Quantum algorithms, like their
classical analogues, can be written in a number of nearly equivalent
ways.  While a classical program is typically composed of a series of
simple boolean functions, such as {\sc nand} and {\sc fanout}, a
quantum algorithm is typically written as a 
product of unitary gates, such as the Hadamard transform $H$, the
controlled--{\sc not} ({\sc cnot}), and the
$\pi/8$-gate~$T$~\cite{Nielsen00a}.  For classical computers, a common
problem is that of compiling a program, in which one typically wishes
to express the program in as few elementary operations as possible.
By analogy, we can raise the principal
questions of {\it quantum compiling}: which sets of gates can be
composed to form what sorts of quantum algorithm, how many of them are
necessary, and what efficient algorithms can be devised to express
quantum programs in terms of a particular set of base gates?

Mathematically, a gate on $n$ quantum bits (qubits) is represented by
a unitary transformation on a $2^n$-dimensional vector space.  We will
denote the set of all determinant-one unitary transformations of a
$d$-dimensional vector space by $SU(d)$.  This space is a manifold and
is hence parameterized by a {\em continuum} of real parameters; for
example, the $2\times 2$ unitary transforms
\be\left(\begin{array}{cc} 
e^{i\alpha}\cos\theta & e^{i\beta}\sin\theta \\ 
-e^{-i\beta}\sin\theta & e^{-i\alpha}\cos\theta 
\end{array}\right)
\end{equation}
parameterized by $\alpha,\beta,\theta$ represent the group $SU(2)$ of
valid single qubit gates.

In contrast, digital quantum algorithms compute with only a {\em finite set}
of base gates (such as those mentioned previously: $H$, $T$, and {\sc
cnot}).  This is a reasonable restriction in real circuit implementations,
since the presence of noise reduces the number of reliably distinguishable
gates to a finite subset of the continuous set.  Finite gate sets are also
intrinsic to fault-tolerant quantum computation, the art of constructing
arbitrarily reliable circuits from unreliable
parts.\cite{Preskill98b,Gottesman98a,Boykin99a,Aharonov99a} Thus, in general
we do not desire perfect computational universality, but only the ability to
approximate any quantum algorithm, preferably without using too many more
gates than originally required.

A set of base gates $\Ai \subset SU(d)$ is {\em computationally universal}
if given any gate $U$, we can find a string consisting of gates from $\Ai$
and their inverses, such that the product of the gates in the string
approximates $U$ to arbitrary precision.  Equivalently, $\Ai$ must generate
a dense subgroup of $SU(d)$.

Which sets of base gates are computationally universal?  It turns out
that probabilistically speaking, almost all of them
are~\cite{Lloyd95a,Deutsch95a}.  If base gates are chosen at random,
then all but a set of measure zero are computationally universal.  The
idea is that if the eigenvalues of the base gates have phases that are
irrationally related to $\pi$ (which occurs with probability one),
then taking powers of them allows each base gate to approximate a
one-parameter subgroup to arbitrary precision, just as integer
multiples of a random vector modulo a lattice will almost always fill
space.  Furthermore, the base gates will almost always lie on
different one-parameter subgroups, which will generate all of $SU(d)$
with probability one.

Given that compiling is generically possible, it is vital to determine
{\em how short a string of base gates is typically required to
approximate a given gate to a specified precision}; this is the
question we consider in this paper.  The construction described by
Lloyd~\cite{Lloyd95a} requires using a number of base gates
exponential in $\log 1/\epsilon$ to achieve a precision of $\epsilon$.
This is an unreasonable cost for many applications.  However,
Solovay~\cite{Sol} and Kitaev~\cite{Kitaev97b} have independently
described an efficient (meaning its running time is polynomial in
$\log 1/\epsilon$) algorithm for quantum compiling that produces
strings of length only $O(\log^c(1/\epsilon))$, where $c$ is a
constant between 3 and 4. \cite{Harrow01a} The algorithm works by
constructing successively finer $\epsilon$-nets; finite sets of gates
that can approximate any element of $SU(d)$ to an accuracy of
$\epsilon$.

On the other hand, as we will later discuss, since a ball of radius
$\epsilon$ in $SU(d)$ has volume proportional to $\epsilon^{d^2-1}$,
it takes $O((1/\epsilon)^{d^2-1})$ different strings of gates to
approximate every element of $SU(d)$ to a precision of $\epsilon$.
Therefore, no algorithm will ever be able to reduce $c$ below 1.
Furthermore, it is unlikely that the successive approximation method
used by the Solovay-Kitaev theorem will be able to do better
than $c=2$.\cite{Harrow01a} This still leaves open the question of
whether some other technique could establish an upper bound
asymptotically smaller than the one achieved by the Solovay-Kitaev
theorem.

Here, our main result is that for at least some univeral sets of base
gates only $O(\log 1/\epsilon)$ gates are sufficient to approximate
any gate to a precision $\epsilon$ (i.e. $c=1$).  This is within a
constant factor of the lower bound obtained from counting arguments.
We say that these base gates are not only computationally universal,
but also {\em efficiently universal}, since using them for quantum
compiling requires a string length that is optimal up to a constant
multiplicative factor.

We present this result as follows.  The set of strings from a fixed
computationally universal set of base gates cover $SU(d)$ increasingly
densely and uniformly, as the string length grows.\cite{Arnold62}
First, in Section~\ref{sec:prelim}, we quantify how quickly this
occurs by introducing a framework for comparing the distribution of
strings with the uniform distribution.  We use this formalism in
Section~\ref{sec:efficient} to identify a condition on base sets that
implies their efficient universality.  In Section~\ref{sec:some} we
then combine this condition with results from the literature to show
that efficiently universal gate sets for Hilbert spaces of any finite
dimension.  Section~\ref{sec:lowerb} discusses lower bounds for
compilation and demonstrates the optimality of the result; we conclude
with open questions and further directions.

%%%%%%%%%%%%%%%%%%%%%%%%%%%%%%%%%%%%%%%%%%%%%%%%%%%%%%%%%%%%%%%%%%%%%%%%%%%%%
\section{Preliminaries}\label{sec:prelim}

We begin by developing a metric of how well strings drawn from a 
finite set of gates approximate arbitrary elements of $SU(d)$. 
 
Let $dg$ be the Haar measure on $SU(d)$ normalized so that $\int
dg=1$.  Consider the Hilbert space $L^2(SU(d))$ with norm defined by
the usual inner product $\langle \psi,\varphi\rangle\equiv
\int\psi(g)^*\varphi(g)dg$.  
The norm of a linear transformation on $L^2(SU(d))$ is given by
\be
	|M|\equiv\sup\left\{\|Mf\|\big\arrowvert 
	f\in L^2(SU(d)),\|f\|=1\right\}
\,.
\end{equation}
When $M$ is bounded and hermitian, the norm is simply the supremum of
its spectrum and as a result, $|M^n|=|M|^n$.

Define a representation $U\mapsto \tilde{U}$ of $SU(d)$ on
$L^2(SU(d))$ by 
\be 
	\tilde{U}f(x)=f(U^{-1}x) 
\,.  
\end{equation} 
Using the right invariance of the Haar measure, we see that
$\tilde{U}$ is unitary.  For any finite set $\Ai \subset SU(d)$,
define the mixing operator $T(\Ai)$ by 
\be 
	T(\Ai) = \frac{1}{2|\Ai|}\sum_{A\in \Ai} \tilde{A} + 
	\tilde{A}^{-1} 
\,.  
\end{equation} 
All such $T$ are hermitian and have norm one.  We will often simply
write $T$ instead of $T(\Ai)$.  These represent averaging the action
of the elements of $\Ai$ and their inverses on a function; when the
function is a probability distribution on $SU(d)$ we can think of $T$
as multiplying by a random element of $\Ai$.

Applying $T^n$ represents averaging over the action of words of length
$n$.  Denote the set of words of length $n$ made up of elements of
$\Ai$ and their inverses by $W_n( \Ai )$, or when the set $\Ai$ is
understood, simply $W_n$.  This set comprises $(2|\Ai|)^n$
words, though as matrices there are generally some duplicates since
substrings such as $AA^{-1}=\mathbf{1}$ for all $A\in\Ai$. For any
positive integer $n$, expanding $T^n$ gives
\be 
	T^n = \sum_{w\in W_n} \frac{\tilde{w}}{(2|\Ai|)^n}
\,.
\end{equation}

We want to compare $T^n$ to the integral operator $P$.
\be 
	Pf(h)=\int f(gh) dg = \int f(g) dg
\,.
\end{equation}
Note that $P$ is the projection operator onto the set of constant
functions on $SU(d)$, and hence $P=P^\dag$ and $P^2=P$.  
It is not hard to show that $TP=P=PT$ and consequently
\be 
	(T-P)^n=T^n-P \label{eq:TP-identity}
\,.
\end{equation}

The metric for comparing $T(\Ai)$ to $P$ is given by
\be
	\Lambda(\Ai) \equiv |T(\Ai)-P|
\,.
\end{equation}
From Eq.~\ref{eq:TP-identity} and the hermiticity of $T$ and $P$, it
follows that
\be
	\Lambda(\Ai)^n = |T^n(\Ai)-P|
\,.
\end{equation}
If one thinks of $T^n$ as a Riemann sum then $\Lambda$ serves as to
quantify how quickly $T^n$ converges to the integral.  It has been
shown \cite{Arnold62} that if $\Ai$ is a computationally universal set
that all the eigenvalues of $T-P$ have absolute value strictly less
than one.  However, this only implies that $\Lambda(\Ai)\leq 1$, since
$T-P$ has an infinite number of eigenvalues.

The proof of main result of our paper---that efficiently universal sets of
gates exist---is divided in the next two sections.  In
Section~\ref{sec:efficient} we show that $\Lambda(\Ai)<1$ implies that
$\Ai$ is efficiently universal and in Section~\ref{sec:some} we
demonstrate that for any $d$ an efficiently universal set of gates can
be found in $SU(d)$.

%%%%%%%%%%%%%%%%%%%%%%%%%%%%%%%%%%%%%%%%%%%%%%%%%%%%%%%%%%%%%%%%%%%%%%%%%%%%%
\section{A condition for efficient universality}\label{sec:efficient}

\begin{theorem}\label{thm:eff-univ}
For any $\Ai\subset SU(d)$ such that $\Lambda(\Ai)<1$, $\Ai$ is
efficiently universal.  Specifically, there exists a constant $C$ such
that for all $U\in SU(d)$, $\epsilon>0$, and $n>C\log 1/\epsilon$,
there is a $w\in W_n(\Ai)$ such that $|w-U|<\epsilon$.
\end{theorem}

Before proving the theorem, we will need to note a fact about the
geometry of $SU(d)$.  For any $d$ and $r_0$, if
$V(r)$ is the Haar measure of a ball of radius $r$ in $SU(d)$, then there
exist constants $k_1$ and $k_2$ such that
\be 
	k_1r^{d^2-1}<V(r)<k_2r^{d^2-1}
\label{eq:ball}
\,.
\end{equation}
for all $r\in(0,r_0)$.  This is true because $SU(d)$
is a $d^2-1$-dimensional manifold and because $V(r)$ does not depend
on the center of the ball under the Haar measure.

Now we can proceed with the proof of Theorem~\ref{thm:eff-univ}:

\begin{proof}
Define $\chi\in L^2(SU(d))$ by
\be
	\chi(g)=\left\{\begin{array}{ll}1 & \mbox{for }|g-I|<\epsilon/2\\
	0 & \mbox{otherwise}\end{array}\right.
\,.
\end{equation} 
Let $V=\|P\chi\|=\|\chi\|^2$ be the measure of the ball around the
identity of radius $\epsilon/2$.  We will not perform this integration,
but recall from Eq.~\ref{eq:ball} that $V>k_1(\epsilon/2)^{d^2-1}$.

Let $T=T(\Ai)$ and $\Lambda=\Lambda(\Ai)$.

First we use the Cauchy-Schwartz inequality to give 
\bea 
&& \left|\left\langle \chi,
(T^n-P)\tilde{U}\chi\right\rangle \right| \leq
\|\chi\|\|(T^n-P)\tilde{U}\chi\|  \n\\
&& \leq \|\chi\|^2|(T^n-P)\tilde{U}| < \Lambda^nV
\label{eq:CS1}
\,.
\eea
Another way to compute the same inner product is
\be
\left\langle\chi,(T^n-P)\tilde{U}\chi\right\rangle
=\left\langle\chi,T^n\tilde{U}\chi\right\rangle - V^2
\label{eq:CS2}
\,.
\end{equation}

 Combining Eq.~\ref{eq:CS1} and Eq.~\ref{eq:CS2} gives that
$\left|\langle\chi, T^n\tilde{U} \chi\rangle - V^2\right| <
\Lambda^n V$.  This means that there exists $C$ which depends only on
$\Ai$ such that if $n>C \log 1/\epsilon$ then $\Lambda^n<V$
and $\langle\chi,T^n\tilde{U}\chi\rangle>0$.  Specifically, it
suffices to choose
\be 
	n>\frac{d^2-1}{\log(1/\Lambda)}\log(1/\epsilon) + 
	\frac{\log(2^{d^2-1}/k_1)}{\log(1/\Lambda)}
\,.
\end{equation}
When this occurs it means that 
\be
\int\chi(g)\sum_{w\in W_n}\frac{\chi(wU^{-1}g)}{(2|\Ai|)^n}dg > 0
\, ,
\end{equation}
which implies that $\exists g\in SU(d)$ and $w\in W_n$ such that
$\chi(g)\neq 0$ and $\chi(wU^{-1}g)\neq 0$.  Thus $|g-I|<\epsilon/2$
and $|wU^{-1}g-I|<\epsilon/2$, implying that $|w-g^{-1}U|<\epsilon/2$.
Combining these and using the triangle inequality gives
$|w-U|<\epsilon$.
\end{proof}

%%%%%%%%%%%%%%%%%%%%%%%%%%%%%%%%%%%%%%%%%%%%%%%%%%%%%%%%%%%%%%%%%%%%%%%%%%%%%

\section{A class of efficiently universal gate sets}\label{sec:some}

In this section we show that for each $d$ there exists a set of gates
$\LPS_d$ in $SU(d)$ such that $\Lambda(\LPS_d)<1$ (and thus $\LPS_d$
is efficiently universal).  We begin with a result
demonstrating this for $SU(2)$ and then extend it to $SU(d)$.

\begin{lemma}[Lubotsky, Phillips and Sarnak]\label{lemma:LPS}
Let 
\bea
V_1&=&\frac{1}{\sqrt{5}}\left(\begin{array}{cc}1 & 2i \\ 2i & 1
\end{array}\right),\,
V_2=\frac{1}{\sqrt{5}}\left(\begin{array}{cc}1 & 2 \\ -2 & 1
\end{array}\right)\n\\
\mbox{ and }V_3&=&\frac{1}{\sqrt{5}}\left(\begin{array}{cc}1+2i & 0\\ 0&1-2i
\end{array}\right).
\label{eq:su2-base-states}\eea Then
$\lambda=\Lambda(\{V_1,V_2,V_3\})=\frac{\sqrt{5}}{3}<1$.  Furthermore, for any
$U_1,U_2,U_3\in SU(2)$, $\Lambda(\{U_1,U_2,U_3\})\geq \lambda$.
\end{lemma}

The proof of this Lemma is presented in
\cite{Lubotsky86a,Lubotsky87a}.  Let $\LPS_2=\{V_1,V_2,V_3\}$, as it is
a family of quantum gates from $SU(2)$ for which $\Lambda$ is strictly
less than one.  The optimality of $\Lambda$ for this set is an
interesting aside, but has little bearing on what follows.

Extending the result to $SU(d)$ will require slightly more
effort.  To this end, if $I_k$ denotes the $k\times k$ identity
matrix, then, for any $U\in SU(2)$ and $2\leq j\leq d$, define
$\beta^{(d)}_j(U)$ to be 
\be
	\beta^{(d)}_j(U)= \left(\begin{array}{ccc}
	I_{j-2} & 0 & 0 \\ 0 & U & 0 \\ 0 & 0 & I_{d-j} \end{array}\right)
	\in SU(d).
\end{equation} 
\,
We will typically omit the $^{(d)}$ where it is understood.

\begin{lemma}[Diaconis and Shahshahani]\label{lemma:diaconis}
Let $\{G_j^i\}, 1\leq i<j\leq d$ be a series of ${d\choose 2}$
independent random matrices in $SU(2)$ that are chosen uniformly
according to a Haar measure.  Then
\be \prod_{i=1}^{d-1}\prod_{j=i+1}^d \beta_j(G_j^i)
\label{eq:diaconis}\end{equation}
is uniformly distributed in $SU(d)$.
\end{lemma}

This Lemma is proved in \cite{Diaconis00a}.  It means that if we had
access to random elements of $SU(2)$ that were completely uniformly
distributed, then we could generate uniformly distributed elements of
$SU(d)$.  When the elements of $SU(2)$ are only approximately uniform,
we can bound the distance to uniformity of the words they form by
using what is known as a hybrid argument: \cite{Bernstein97a}

\begin{lemma}[Bernstein and Vazirani]\label{lemma:hybrid}
If $U_1,\ldots,U_m$, $V_1,\ldots,V_m$ are linear operators such that
$|U_i|\leq 1$, $|V_i|\leq 1$ and $|U_i-V_i|<\delta$, then $|U_m\cdots
U_2U_1-V_m\cdots V_2V_1|<m\delta$.
\end{lemma}

\begin{proof}
If we replace a single $U_i$ in the product $U_m\cdots U_1$ with the
corresponding $V_i$, then the entire product will still change by less
than $\delta$ since $|AB|\leq |A|\cdot|B|$ for any operators $A,B$.
Thus we can construct a series of $m+1$ 
``hybrid'' operators, which start with $U_1\cdots U_m$, end with
$V_1\cdots V_m$ and are each separated by less than $\delta$.  The
proof follows from the triangle inequality.
\end{proof}

We now combine all of the other results in this section to demonstrate
a set of gates in $SU(d)$ for which $\Lambda$ is strictly less than
one.

\begin{proposition}\label{prop:sun-base-states}
For any $d>2$, define $\LPS_d$ by
\be\label{eq:sun-base-states}
	\LPS_d= \{\beta_j (V) \ | \ 1\leq j \leq (d-1), V\in\LPS_2 \} 
\,.
\end{equation}
Then $\Lambda(\LPS_d)<1$.
\end{proposition}
\begin{proof}

The approach of our proof will be to approximate the uniform
distribution in Lemma \ref{lemma:diaconis}, and then we show that this
forces $\Lambda$ to be less than one.  To this end, let $R_m \subset
W_{m{d\choose 2}}(\LPS_d)$ be the set of all products of the form
\be
	\prod_{i=1}^{d-1}\prod_{j=i+1}^d \beta_j(G_j^i)
\end{equation}
such that the $G^i_j$ are selected from $W_m(\LPS_2)$.

From Lemma~\ref{lemma:LPS} we have that $\forall m$,
$|T(V_1,V_2,V_3)-P|^m=\lambda^m$ for some $\lambda<1$.  There are
$\binom{d}{2}$ terms in Eq.~\ref{eq:diaconis}, each of which is
approximated to within an accuracy of $\lambda^m$ by the appropriate
length $m$ substring of $R_m$.  Thus, using the hybrid argument and
Lemma \ref{lemma:diaconis} gives that
\be
	\Lambda(R_m) = \left|\sum_{w\in R_m}\frac{\tilde{w}-P}
	{|R_m|}\right|\leq{d\choose 2}\lambda^m
\,.
\end{equation}
Now, if we let $R'_m$ denote $W_{m{d\choose 2}}-R_m$ then
\bea
\Lambda(W_{m{d\choose 2}}) &\leq&
 \frac{|R'_m|}{|W|} \Lambda(R'_m)
 + \frac{|R_m| }{|W|} \Lambda(R_m) \n\\
&\leq& \left(1-\frac{|R_m|}{|W|}\right) 
+ \frac{|R_m| }{|W|} \Lambda(R_m)\n\\
&=& 1-\frac{|R_m|}{|W|}(1-\Lambda(R_m))
\,.
\eea
If we choose $m$ large enough so that ${d\choose 2}\lambda^m<1$, then this
last expression will be less than one, and $\Lambda(\LPS_d)<1$.
\end{proof}

Thus, efficient quantum compiling is possible for $d$-dimensional
systems, given the appropriate choice of base gate set.

%%%%%%%%%%%%%%%%%%%%%%%%%%%%%%%%%%%%%%%%%%%%%%%%%%%%%%%%%%%%%%%%%%%%%%%%%%%%%
\section{Lower Bounds}\label{sec:lowerb}

This proves that sets of base gates exist which can achieve a
precision of $\epsilon$ in $O(\log 1/\epsilon)$ gates, but can we do
any better?  An $\epsilon$-ball in $SU(d)$ has measure of order
$\epsilon^{d^2-1}$, so if we expect to cover all of $SU(d)$ with
strings of length $n$, then we will require
$(2|\Ai|)^nk_2\epsilon^{d^2-1}>1$, or equivalently,
\be 
	n\geq\frac{d^2-1}{\log 2|\Ai|}\log 1/\epsilon -
	\frac{\log k_2}{\log 2|\Ai|}
\,.
\end{equation}  
Thus the result is optimal up to a constant factor.  This fact is
quite general, since it follows from simple counting arguments.
However, if the assumptions of the problem are relaxed to allow many
gates to act in parallel, then using ancilla qubits it is possible to
approximate single-qubit gates with a circuit of {\em size}
poly($\log 1/\epsilon$) but {\em depth} of only poly($\log \log
1/\epsilon$).\cite{Kitaev02}  This construction, like the one in this
paper, relies on having access to a specific set of base gates; to
date, only the Solovay-Kitaev theorem applies to any computationally
universal set.

In our original problem, though, eliminating the constant linear
factor turns out to be impossible.  Consider any set $\mathcal{A}$ of
$l$ base gates that is {\em not} computationally universal.  Let
$B(\mathcal{A},\delta)$ be the set of gates obtained by perturbing
each gate in $\mathcal{A}$ by no more than $\delta$.  Then
$B(\mathcal{A},\delta)$ has non-zero measure (in $SU(d)^l$), almost
all of its elements are computationally universal and from the hybrid
argument, any string of length $n$ drawn from gates in
$B(\mathcal{A},\delta)$ will be within $n\delta$ of something in the
(non-dense) group generated by $\mathcal{A}$.  Since we can make
$\delta$ arbitrarily small, any fixed prefactor in front of $\log
1/\epsilon$ will fail on a computationally universal set of non-zero
measure for some values of $\epsilon$.

Note that unlike most results about quantum compiling,
this argument also holds if the base gates are parameterized;
say, $A_1,\ldots,A_l$ are elements of the algebra $su(d)$ and a single
operation now has the form $e^{\pm A_it}$, for any $t>0$.  
The above proof demonstrates that there exist sets with non-zero
measure which require arbitrarily many steps, even if the steps are
continuous.  If we measure cost not in terms of number of steps, but
by the total time taken, then we have to modify the argument
slightly.  For small values of $t$, $|e^{A_it}-I|$ is on the
order of $t\delta$, but for large $t$ the difference never gets any
higher than $\delta$.  This means that no matter how many steps we
take, in time $t$, we will stay within $t\delta$ of some non-dense
subgroup and the same result holds.

These results can be obtained more simply by considering the (non-zero
measure) set of gates which are very close to the identity.  If every
gate does very little, then we will need a large number them in order
to accomplish anything.  The reason why universal sets that are very
close to non-universal sets are interesting is because of their
frequent appearance in actual physical systems, such as NMR under the
weak coupling approximation.\cite{Ernst97}

%%%%%%%%%%%%%%%%%%%%%%%%%%%%%%%%%%%%%%%%%%%%%%%%%%%%%%%%%%%%%%%%%%%%%%%%%%%%%
\section{Conclusions}\label{concl}

We have found a condition that implies the efficient universality of a
set of gates and demonstrated a family of gate sets in $SU(d)$ that
satisfy this condition.  This means that given access to such a gate
set, arbitrary quantum gates can be approximated to accuracy
$\epsilon$ using only $O(\log 1/\epsilon)$ gates.  Such knowledge will
likely be invaluable in crafting future physical implementations of
quantum information processing systems.

Many open questions remain, however.  For example, determining or
bounding $\Lambda$ (even numerically) for a given set of base gates
seems to be very difficult, though it is likely an important step in
determining the prefactor $C$, which measures how effective a set of
gates would be for compiling.  The method used by \cite{Lubotsky86a,
Lubotsky87a} involves specialized arguments from number theory that do
not generalize easily to other sets of gates or to $SU(d)$ for $d>2$.
Our proof (like the Solovay-Kitaev algorithm) also requires the
ability to perform the inverse of each gate in the base set.  This
restriction feels unnecessary, yet very little is known in the
case where inverses are unavailable.

More broadly, it is also generally unknown which gate sets are
efficiently universal and when $\Lambda<1$.  Note that
$\Lambda(\Ai)<1$ implies that $\Ai$ is efficiently universal, but the
converse is not known to be true.  Thus it is possible that the
questions of efficient universality and $\Lambda$ being less than one
will be settled separately.

However, if $\Lambda(\Ai)$ were to be a continuous function of $\Ai$
(for fixed $|\Ai|$), then the situation would simplify considerably.
In this case, it is not hard to show that $\Lambda(\Ai)<1$ if and only
if $\Ai$ is computationally universal, so that computational
universality, efficient universality and $\Lambda<1$ would all become
equivalent conditions.  We suspect that this is the case, but have
been unable to prove it.

Finally, the techniques used in our results do not suggest any
efficient (i.e. running time polynomial in $\log 1/\epsilon$)
algorithms for quantum compiling.  The most important, and possibly
most difficult, open problem remaining is to find a polynomial time
algorithm to approximate any unitary gate by a fixed efficiently
universal set of base gates with a string whose length saturates the
$O(\log 1/\epsilon)$ bound.

We thank Persi Diaconis, Michael Freedman, Neil Gershenfeld, David
Jerison, Seth Lloyd and Michael Nielsen for useful conversations and
assistance, and are particularly grateful to Alexei Kitaev for
pointing out a flaw in an earlier version of this paper and sharing
with us a draft of \cite{Kitaev02}.  AWH acknowledges support from the
Army Research Office under the SUSPENSE program.  ILC and BHR are
supported by the Things That Think consortium.  This work was also
supported by the DARPA QuIST project on Quantum Architectures.

%%%%%%%%%%%%%%%%%%%%%%%%%%%%%%%%%%%%%%%%%%%%%%%%%%%%%%%%%%%%%%%%%%%%%%%%%%%%%
% References
 
\bibliographystyle{prsty}
\bibliography{ub}

\end{document}